\documentclass{hep99}
\usepackage{epsfig}
\begin{document}
\title{First Electro-Optical Detection of Charged Particles}
\author{D.M. Lazarus$^a$, V. Castillo$^a$, L. Kowalski $^b$, D.E. 
Kraus$^c$, R. Larsen$^a$, B. Magurno$^a$, D. Nikas$^a$, C. Ozben$^a$, 
Y.K. Semertzidis$^a$, T. Srinivasan-Rao$^a$, T. Tsang$^a$}
\address{$^a$ Brookhaven National Laboratory, Upton, NY 11973, U.S.A.\\
$^b$ Montclair State University, Upper Montclair, NJ 07043 U.S.A.\\
$^c$ University of Pittsburgh, Pittsburgh PA, U.S.A.}
\abstract{We have made the first observation of a charged particle beam by means
of its electro-optical effect on the polarization of laser light in a
birefringent crystal. The modulation of the laser light during the passage 
of a pulsed electron beam was observed using a fast photodiode and a digital
oscilloscope. The fastest rise time measured in a single shot, 120 ps, was 
limited by the bandwidth of the oscilloscope and the associated electronics. 
This technology holds promise for detectors of greatly improved spatial and 
temporal resolution for single relativistic charged particles as well as 
particle beams.}
\maketitle
\fntext{\dag}{Supported in part by the U.S. Department of Energy under 
Contract No. 
DE-AC02-98CH10886.}
An effort has been initiated to develop an ultra-fast charged particle detector 
based on the birefringence, and hence the phase difference induced between 
orthogonal components of polarization (ellipticity), in an optical medium by 
the electric field of a relativistic charged particle~\cite{gls}. The 
electro-optical effect in amorphous optical media is known as the Kerr 
effect~\cite{kerr} and is quadratic in the electric field $E$.
In uniaxial crystals the induced ellipticity is linear in the  E-field and 
is known as the Pockels effect~\cite{yariv}. The induced phase delay can then 
be given by $\phi = \pi (V/V_\pi)$ with $V$ the applied voltage and $V_\pi$ 
the voltage required for producing a phase shift between orthogonal components
of polarization equal to $\pi$ radians or an ellipticity of $\pi/\sqrt{2}$ 
since the maximum ellipticity is induced when the polarization is at 
$45^\circ$ with respect to the applied electric field direction.

The passage of a tightly bunched relativistic electron beam within several
millimeters of a commercially available LiNbO3 crystal~\cite{utp} coupled to 
polarization maintaining fibers of 4 microns diameter propagating
polarized infrared light ($\lambda = 1.32 \mu {\rm m}$) was observed by the 
effect of a microbunched electron beam on the polarization of the light in the 
crystal. This was determined by means of a $\lambda/4 $ plate which 
converted the induced ellipticity to a rotation of the initial linear 
polarization which had previously been nearly extinguished by the analyzer
(crossed polarizer). 
The resulting modulation of the transmitted laser light was detected by a 
photodiode of 45 GHz bandwidth and pre-amplifier. The output 
signals were digitized in a 7 GHz sampling oscilloscope and stored 
in memory. The optics setup is indicated in Fig.1
\begin{figure}
\begin{center}
\vspace*{.3cm}
\epsfig{file=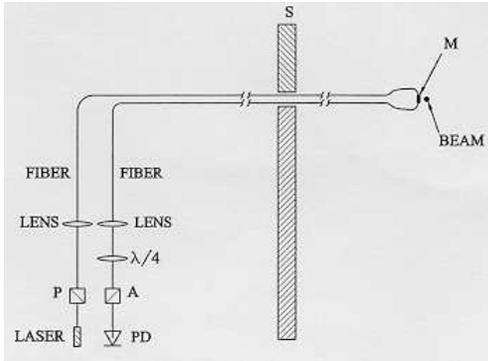, width=65mm}
\end{center}
\caption{The experimental setup for detecting a charged particle beam. 
The LiNbO$_3$ crystal was located at the beam position indicated by  E-field.
The positions of the polarizer (P), lenses (L), analyzer (A) and photodiode 
detector(D) are schematically indicated.}
\end{figure}


A bunched charged particle beam creates an electric field, $E$, at a distance 
$r$ that is well approximated by Coulomb's Law multiplied by $\gamma$, the 
relativistic Lorentz factor, and $N_e$ the number of particles in the beam bunch
for a minimum distance $r_0\gg$ than the dimensions of the beam bunch.
The LiNbO$_3$ crystal had $V_{\pi}=5$~V with an electrode separation of 
$15\mu$m and a length of $l=1.5$~cm. The integral 
$\int E\,dl=\int 5V/(15\times 10^{-6}\,{\rm m})\,dl=5000$~V produces 
$\pi/\sqrt{2}$ radians maximum ellipticity.

The integral of the electric field of the particle bunch over the crystal 
dimensions for a beam bunch located at the mid-plane orthogonal to the length 
of the crystal at a distance $r_0$ is
\begin{equation}
\int Edl=\frac{N_e\gamma q}{2\pi\epsilon_0 r_0}(1-\sin\theta)
\end{equation}
where $\theta$ is the angle subtended by the crystal length, $l$ and the 
direction to the beam center from the end of the crystal. For $r_0=5$~mm,
$\int Edl=2.6\times 10^{-7}\gamma N_e {\rm V}$ which should yield an 
ellipticity of $\phi=\gamma N_e\times 0.1~{\rm nradian}$. The signal to noise 
ratio for a measurement of this type that is photon statistics limited is 
given by 
\begin{equation}
SNR = \phi \sqrt{P T q_p \over 2 h\nu} 
\end{equation}
where $P$ is the laser power, $T$ the time resolution or inverse of the 
detection system bandwidth, $q_p$ the quantum efficiency of the photodiode, 
and $h\nu$ the laser photon energy. 

A 45 MeV kinetic energy electron beam at the Brookhaven National Laboratory 
Accelerator Test Facility containing up to 1 nC in a diameter of approximately 
1 mm with 10 ps duration and a repetition rate of 1.5~Hz was scanned
across the crystal. Measurements were made in the single shot mode. 
The extinction was generally close to $10^{-3}$.

With $\gamma=88$, $P=10$~mW, $T=100$~ps and $q_p=0.8$ and $h\nu=0.9$~eV, Eq.2
gives the required number of electrons in the beam for $SNR=1$, 
$N_e\sim 6.8\times 10^4$ for detection of a single beam bunch. However, the
7 GHz bandwidth limits the sensitivity to the inherently much faster signal
by as much as factor of five implying $\approx 3\times 10^5$ singly charged 
particles are required for detection of a single ATF beam bunch. The ATF beam 
of $N_e=6.3\times 10^9$ per bunch was sufficient to generate a detectable 
signal.
The maximum modulation of the light intensity was about $9\%$ of its DC level.
\begin{figure}
\begin{center}
\vspace*{-.9cm}
\epsfig{file=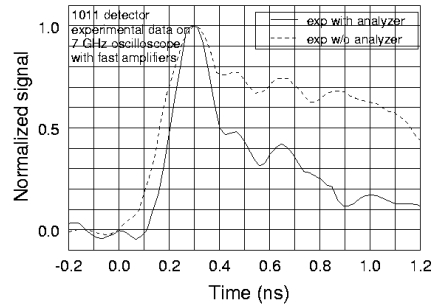, width=60mm,angle=-90}
\end{center}
\vspace*{-1.1cm}
\caption{The polarization dependent signal (solid line).  The electron beam 
was $\approx 0.5$~cm from the crystal.  The polarization independent
signal is indicated by the dashed line.}
\end{figure}

The polarization dependent signal is displayed in Fig. 2 (solid curve). A 
signal was also obtained when the crystal intercepted the beam (dashed curve)
which was found to be independent of the analyzer and hence is polarization 
independent. It does not change sign with polarization orientation and it has a 
significantly longer decay time constant than the polarization dependent signal.

In order to achieve single particle detection with enhanced time resolution, 
a high power pulsed laser of $10^8$~W, e.i. 10mJ for 10 ps and a transient 
digitizer of greater bandwidth, e.g. 100 GHz would require 
$\gamma N_e\approx 20$ for single charged particle detection with a $SNR=1$. 
The use of a high power pulsed laser would preclude the detection of randomly 
occurring events because of the inability to trigger the laser. However, this
would not be a problem at colliding beams machines where the beams are bunched 
to maximize the luminosity. Particles of $\gamma Z>20$ would be detectable at
RHIC and the LHC.

A detector constructed of parallel rows of electro-optical crystals with a 
separation of $100\mu$~m
would easily provide single particle detection in LiNiO3 based detectors. 
However, the small size and high cost of these crystals would severely limit 
their applicability.

We have recently produced poled optical fibers~\cite{pole}. If they or other 
moderately priced electro-optical materials can yield adequate sensitivity 
for single particle detection, a new generation of ultrafast fine grained 
charged particle detectors may be at hand.

\end{document}